\documentstyle[epsf,graphicx]{mn}

\def\x2{$\chi^{2}$}

\def\asca{{\it ASCA }}
\def\rosat{{\it ROSAT }}

\def\einstein{{\it EINSTEIN }}

\def\x2{$\chi^{2}$}
\def\lunits{$\rm{erg\,s^{-1}}$}

\def\junits{$h~\rm{erg~s^{-1}~Mpc^{-3}}$}
\newcommand{\mincir}{\raise
-2.truept\hbox{\rlap{\hbox{$\sim$}}\raise5.truept\hbox{$<$}\ }}
\newcommand{\magcir}{\raise
-2.truept\hbox{\rlap{\hbox{$\sim$}}\raise5.truept\hbox{$>$}\ }}
\newcommand{\minmag}{\raise
-2.truept\hbox{\rlap{\hbox{$<$}}\raise6.truept\hbox{$<$}\ }}
\newbox\grsign \setbox\grsign=\hbox{$>$} \newdimen\grdimen \grdimen=\ht\grsign
\newbox\simlessbox \newbox\simgreatbox \newbox\simpropbox
\setbox\simgreatbox=\hbox{\raise.5ex\hbox{$>$}\llap
     {\lower.5ex\hbox{$\sim$}}}\ht1=\grdimen\dp1=0pt
\setbox\simlessbox=\hbox{\raise.5ex\hbox{$<$}\llap
     {\lower.5ex\hbox{$\sim$}}}\ht2=\grdimen\dp2=0pt
\setbox\simpropbox=\hbox{\raise.5ex\hbox{$\propto$}\llap
     {\lower.5ex\hbox{$\sim$}}}\ht2=\grdimen\dp2=0pt

\begin{document}

\title[The X-ray luminosity function  of  galaxies] {
The X-ray luminosity function of local galaxies}

\author[I. Georgantopoulos et al.]
       {I. Georgantopoulos$^{1}$, S. Basilakos$^{1,2}$, M. Plionis$^{1}$\\
$^{1}$ National Observatory of Athens, I. Metaxa \& B. Pavlou, Lofos Koufou, 
 Palaia Penteli, 
15236, Athens, Greece \\
$^{2}$ Physics Dept, University of Athens, Panepistimiopolis, 15783, Athens, Greece} 

\maketitle

\label{firstpage}

\begin{abstract}

We present an estimate of the local X-ray luminosity function and emissivity 
 for different subsamples of galaxies namely Seyferts, LINERS, star-forming
 and  passive (no-emission-line) galaxies. 
 This is performed by convolving their optical 
 luminosity function, as derived from the Ho et al. 
 spectroscopic sample of nearby galaxies with the 
corresponding $L_x/L_B$ relation. 
 The local galaxy emissivity is $\approx 1.6\times 10^{39}$ \junits 
 in  agreement with the results from a number of 
 cross-correlation analyses using large-area surveys. 
  From our analysis, it becomes  evident that the largest fraction of the 
 galaxy emissivity comes 
  from galaxies associated with AGN (Seyferts but also LINERS) 
 while  the contribution of star-forming and passive 
 galaxies is small. This independently supports the view that most of the 
 yet unidentified X-ray sources in deep \rosat fields 
 which are associated with 
 faint optical galaxies, do harbour an AGN.

\end{abstract}

\begin{keywords}
galaxies: starburst - galaxies:evolution -quasars:general 
 -diffuse radiation - X-rays:galaxies - X-rays:general.  
\end{keywords}

\section{INTRODUCTION}
 The launch of the X-ray satellite \rosat  has brought 
 great progress in understanding the origin of the 
 X-ray background (XRB) (for a review see Fabian \& Barcons 1992). 
 Deep X-ray surveys have resolved about 70 per cent of the 
 XRB at soft energies (0.5-2 keV) (Hasinger et al. 1998). 
 Spectroscopic follow-up observations have demonstrated that the
 majority of these sources are broad-line, luminous AGN (QSOs) eg. 
 Shanks et al. (1991), Georgantopoulos et al. (1996), McHardy et al. 
 (1998), Schmidt et al. (1998). 
 The QSO luminosity function (LF) derived on the basis 
 of these surveys (Boyle et al. 1993, Boyle et al. 1994,
  Page et al. 1996, Jones et al. 1997) suggests that 
 QSOs cannot contribute the bulk of the XRB at these energies. 
 Indeed,  at faint X-ray fluxes there are many sources 
  associated with optical galaxies which present 
 narrow emission lines  (Boyle et al. 1995, 
 Griffiths et al. 1996, McHardy et al. 1998).  
 Roche et al. (1995, 1996), Almaini et al. (1997) and Soltan et al. (1997) 
 quantified the contribution of these  NELGs
 by cross-correlating optically selected galaxies with  ROSAT 
  XRB fluctuations. They found that NELGs can easily contribute the bulk 
  of the XRB together with QSOs at soft energies.
  Similar attempts were made by Refregier, Helfand \& McMahon (1997) 
 using \einstein data and also in the hard band (2-10 keV) by 
 Lahav et al. (1993), Miyaji et al. (1994) and Carrera et al. (1995) 
  who cross-correlated nearby {\it IRAS} galaxy catalogues with
 {\it HEAO-1} or {\it Ginga} data.  Although there is mounting 
 evidence that most of these galaxies 
 do present either high excitation lines or broad wings 
 and thus they are associated with AGN activity 
 (eg Schmidt et al. 1998), their exact nature 
  and contribution to the XRB remains unclear. 
 
 Here, we provide an independent estimate of the 
 local galaxy X-ray LF and emissivity.
 More importantly, we assess the  contribution of 
 different classes of galaxies (Seyferts, LINERS, star-forming and
  passive galaxies) to the galaxy X-ray emissivity. 
 This is done by convolving the optical LF  
 as derived from the  Ho et al. (1995) sample of nearby 
 galaxies with the corresponding $L_x/L_B$ relation (section 2).
 The same method has been employed using {\it IRAS} data 
 (Griffiths \& Padovani 1990, Treyer et al. 1992, Barcons et al. 1995).
  Schmidt, Boller \& Voges (1996) also present a preliminary analysis 
 of the local galaxy luminosity function using \rosat all-sky survey data. 
  However, the use of the Ho et al. sample presents the advantage that 
  excellent quality nuclear spectra exist for each galaxy;
 thus it provides us with one of the less biased 
 samples against low-luminosity AGN. 
 Moreover, the use of an optical sample may introduce less bias against 
  passive galaxies 
 and LINERS most of which are associated with early-type 
 galaxies and thus are probably 
 under-represented in IR selected samples. 
  Our findings are compared with both the 
 cross-correlation  (eg Soltan et al. 1997) 
 as well as  the {\it IRAS} LF (eg Barcons et al. 1995)
 results  in section 3. 

\section{Method and results}

The X-ray galaxy LF can be obtained 
 by convolving the optical LF 
 with the $L_x/L_B$ relation (see Avni \& Tananbaum 1986) 
 ie. by deriving the bivariate optical/X-ray LF:

\begin{displaymath}
 \Phi(l_x)= \int \Phi(l_B) ~ \phi(l_x|l_B) dl_B
\end{displaymath}
where $l_x$ and $l_B$ denote the logarithms of the X-ray (0.2-4 keV) 
and the optical (B) luminosity respectively; 
 $\Phi(l_B)$ is the optical  luminosity 
 function per unit logarithmic luminosity interval 
 and  $\phi(l_x|l_B)$, 
 is the conditional probability function ie it gives the 
 distribution of $l_x$ around the average 
 X-ray luminosity $\langle l_x \rangle$ 
 at a given optical luminosity $l_B$.
 We have used $H_o=100 \rm km ~s^{-1}~Mpc^{-1}$ 
 and $q_o=0.5$ throughout the paper.

\begin{figure}
\mbox{\epsfxsize=8cm \epsfysize=7cm \epsffile{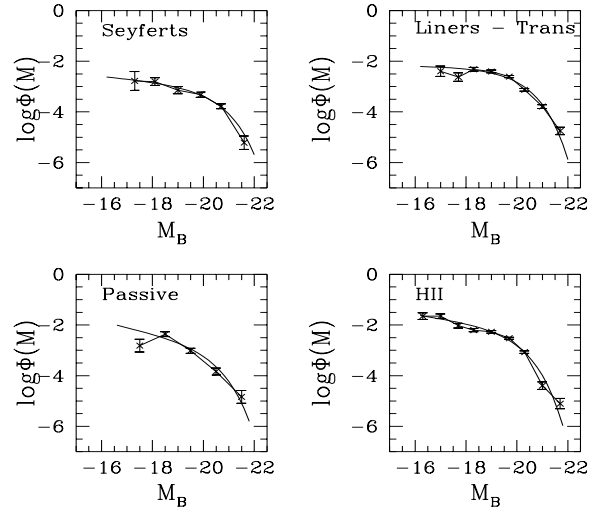}}
\caption{The optical LF of the different sub-samples 
 of galaxies in the Ho et al. sample: Seyferts, LINERS, HII and 
 passive galaxies. The solid line denotes the best-fit Schechter 
 function to the $1/V_{max}$ points (crosses).}
\label{opt}
\end{figure}

We have used the Ho et al. (1995, 1997a) 
 spectroscopic sample of nearby galaxies 
 in order to derive the optical LF for various classes 
 of galaxies. The above is a magnitude limited sample
 ($B<12.5$)  of 486 galaxies above $\delta>0$. 
  Excellent nuclear spectra (high signal-to-noise, medium resolution) 
  and thus bona-fide spectroscopic identifications exist for each galaxy
 (Ho et al. 1997a).  
  14 per cent of the galaxies do not present 
 emission lines in their nuclei 
 and hence can be classified as passive galaxies 
 or ``early-type galaxies'' 
 on the basis of spectroscopy rather than on morphology. 
 At least forty per cent of the galaxies have AGN like spectra: 
   Seyfert, LINERS or transition objects ie.  
 with composite LINER/HII spectra. The remaining 
  nuclei can be classified as star-forming (HII) (Ho et al. 1997b). 
  The use of the above sample offers the possibility to  
 construct a LF for various spectroscopic subsamples of galaxies and 
 therefore to distinguish the  relative contribution of 
 AGN and normal galaxies to the XRB. 
 The magnitudes listed in Ho et al. (1997a) are corrected for 
  both Galactic and intrinsic reddening. 
The distances have been corrected according to the 
 Virgo infall model of Tully \& Shaya (1984). 
 We have used only galaxies with $|b|>20^\circ$ in order to 
 minimize the effects of Galactic reddening.
 We are then left with 416 galaxies with $B<12.5$. 
 We derive the optical LF using the classical $1/V_{max}$ method 
 (Bingelli, Sandage \& Tammann 1988).  
  The $1/V_{max}$ 
 points are then fitted with a Schechter function 
 in order to obtain a parametric expression for the LF: 
\begin{displaymath}
 \Phi(M)=  \phi_\star ~ 10^{[0.4(M_\star-M)(\alpha+1)]} ~ 
 \exp[-10^{0.4(M_\star-M)}] 
\end{displaymath}
  The results are presented in table 1: column (1) gives the 
 type of galaxies; column (2) gives the 
 normalization $\phi_\star$ in units of $\rm Mpc^{-3}~mag^{-1}$;
 columns (3) and (4)  
 give the slope $\alpha$ and characteristic magnitude 
 $M_\star$ of the LF together with the 
 associated 1$\sigma$ error bars; 
 finally column (5) lists the reduced $\chi^2$. 
 Note that the best-fit LF for all galaxies together, 
 yields $\phi_\star=1.8\times10^{-2}$ $\rm Mpc^{-3}~mag^{-1}$, 
 $\alpha=-1.2^{+0.04}_{-0.13}$ 
 and $M_\star=-19.7^{+0.05}_{-0.50}$,  with a  reduced 
 $\chi^2$ of 0.5.  
 The galaxy LF derived by Loveday et al. (1992) from the APM sample 
 yields $\phi_\star=1.4\times10^{-2}$, $\alpha=-1.0$ and 
 $M_\star=-19.5$. Therefore, 
 the Loveday et al. (1992) fit is in good agreement with our results 
 suggesting that our crude approach provides a good   
 approximation to the true galaxy LF.  
  We have also verified that 
 excluding a $10^{\circ}$ region around Virgo does not change our results 
 appreciably.
 In Fig. \ref{opt} we present the LF for the different subsamples 
 of galaxies. Henceforth, we include the ``transition''
 class of objects in the LINERS sample. 

\begin{table}
\caption{The best-fit results for a Schechter optical LF}

\begin{tabular}{ccccc}
Type & $\phi_\star$ &  $\alpha$   &  $M_\star$ & $\chi^2/dof$  \\ \hline 
Seyferts & 0.001 &$-1.2^{+0.3}_{-0.5}$&$-20.1^{+0.3}_{-0.1}$& 0.27\\
LINERS &  0.005& $-1.0^{+0.2}_{-0.3}$ &$-19.7^{+0.2}_{-0.3}$ & 0.73 \\
HII & 0.009 &$ -1.3^{+0.2}_{-0.1}$  & $-19.5^{+0.2}_{-0.4}$ &0.62 \\
passive & 0.002 &$-1.5^{+0.4}_{-0.6}$ &$-19.8^{+0.3}_{-0.4}$&3.0\\
\end{tabular}
\end{table}

Next, we derive the $L_x/L_B$ relation for the above 
subclasses of galaxies. Many galaxies from the Ho et al. sample 
 are listed in the  X-ray catalogue of galaxies  
 of Fabbiano, Kim \& Trinchieri (1992). This catalogue contains 
 \einstein detections and $3\sigma$ upper limits 
 of nearby galaxies in the 0.2-4 keV band. 
 After excluding objects where the X-ray emission may not be 
 associated with the galaxy (see Fabbiano et al. 1992 for details) 
 we are left with 164 objects (of which 69 are upper limits). 
  We have performed a regression analysis ($\log L_x (\rm 0.2-4 keV)$  
 against $\log L_B$) 
 using the EM algorithm of the {\small ASURV} 
  survival analysis package (Isobe, Feigelson \& Nelson 1986). 
 The results are presented in table 2: column 1 
 gives the galaxy type together with the 
 number of galaxies used; columns (2)  (3) and (4) list 
 the slope, intercept and the dispersion $\sigma$ 
 (assuming a Gaussian distribution around the best-fit line) 
 of the $\log L_x$ vs $\log L_B$
 relation while column (5) gives the mean X-ray luminosity. 
 In our analysis there is the inherent assumption that 
 the above 164 galaxies represent an 
 unbiased subsample of the Ho et al. survey. 
 As our sample consists mainly of targets rather than  
 serendipitously observed sources, there may be possible biases
 in the derived $L_x/L_o$ relation. 
 Therefore, we have attempted to test the validity of the 
 ``fair sample'' assumption by dividing our original 
 sample into two subclasses on the basis 
  of the galaxy's off-axis angle, $\theta$: 
  sources with $\theta < 5$ arcmin mostly consisting of targets and objects 
 with $\theta>5$ arcmin which are mostly serendipitous sources. 
  The $L_x/L_o$ results for the above two subsamples are found to be 
  statistically equivalent.  

\begin{table}
\caption{The regression results for
 the $\log L_{\rm 0.2-4 keV}$ vs. $\log L_B$ relation}
\begin{tabular}{ccccc} 

Type & slope & Intercept & dispersion & $\rm \langle l_x \rangle $ \\ \hline 
Seyferts (22) & 1.62 & -30.10 & 0.78 & 41.0 \\
LINERS  (59) & 0.90 & 0.54 & 0.74 & 39.9 \\
HII (54) & 0.89 & 0.65 & 0.47 & 39.3 \\
passive (29) & 1.81 & -39.2 & 0.78 & 39.2 \\ 
\end{tabular}
\end{table}

\begin{figure}
\mbox{\epsfxsize=8cm \epsfysize=7cm \epsffile{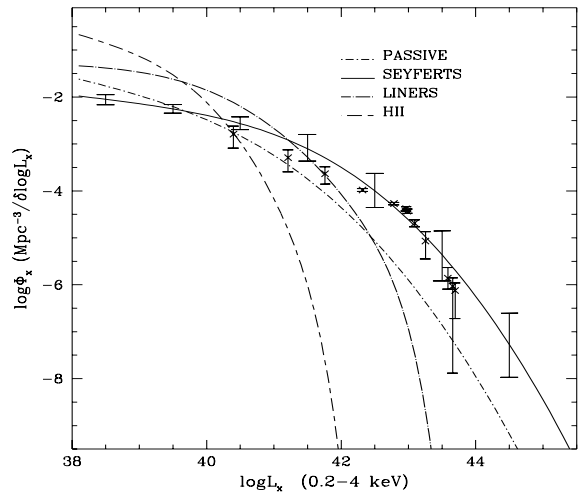}}
\caption{The X-ray LF, for different subsamples of 
galaxies (Seyferts, Liners, star-forming and passive). 
 The $1\sigma$ errors on the Seyfert LF are also shown. 
The crosses  denote the local QSO X-ray LF derived 
from the RIXOS sample (Page et al. 1996)}
\label{xray}
\end{figure}

 The derived X-ray LFs for various subclasses of galaxies are 
 presented in Fig. \ref{xray}. 
 We plot the $1\sigma$ error-bars only for the  
 Seyferts in order to avoid confusion. 
 The errors  were derived by varying the 
 optical LF slope $\alpha$ and the $L_x/L_B$ relation within  
 their 1$\sigma$ confidence limits. 
 For comparison we also plot the {\it RIXOS} QSO X-ray LF from 
 Page et al. (1996). 
 Using the  LF derived above, we estimate the 
 local volume emissivity and the contribution of galaxies 
 to the  0.2-4 keV XRB. The derived contribution 
 sensitively depends on the cut-off redshift, $z_{max}$, 
  the  rate of evolution  
  as well as the  X-ray spectral index  (eg Lahav et al. 1993). 
 Here, we choose $z_{max}=4$. We use a spectral index of 
 $a_x=0.7$ comparable to the X-ray spectrum of NELGs in the 
 ROSAT band (Almaini et al. 1996, Romero-Colmenero et al. 1996). 
  We first adopt a simple no-evolution model.  
 We choose the lower limit of integration in luminosity, $L_{min}$, to be 
 as low as $L_x =10^{38}$ \lunits ie. the luminosity of a solar mass 
  X-ray binary radiating at the Eddington limit. 
 Our results do not critically depend 
  on the lower limit of integration. When we increase the limit of integration 
to $L_x=10^{40}$ \lunits, the Seyfert emissivity reduces by 
  5 per cent. 
 Note, however, that our derived LF depends 
 sensitively on the dispersion of the $L_x/L_o$ relation,
 in the sense that the larger the dispersion the 
 higher the number density of bright objects and thus the emissivity. 
 If the dispersion 
 is systematically over-estimated, then our results should be viewed 
 only as upper limits to the emissivity.
 This may be the case in the Seyfert subsample where the 
 presence of intrinsic absorption especially 
 in low-luminosity objects may broaden the $L_x/L_o$ distribution
 (see Franceshini, Gioia, Maccacaro 1986).  
 The results are presented in table 3: $j$ denotes the emissivity 
 in units of $10^{39}$ \junits   
  while {\it f} denotes the fractional contribution to the 
 0.2-4 keV XRB, assuming no evolution. 
 The observed XRB intensity in the above band 
 was estimated integrating the expression $9\times E^{-0.4}$ 
 $\rm keV~cm^{-2}~s^{-1}~sr^{-1}~keV^{-1}$ (Gendreau et al. 1995).  
  In a strong evolution scenario ie. $j \propto (1+z)^k$ with 
 the evolution parameter  $k=3$, in accordance to the  
 results of Almaini et al. (1997) for the ROSAT NELGs,
 the Seyferts alone would produce 90 per cent of the XRB intensity.

\begin{table}
\caption{The X-ray local galaxy emissivity and 
fractional contribution to the 0.2-4 keV XRB
in the case of no evolution}

\begin{tabular}{cccc}

type & emissivity & fraction \\ 
     & ({\small $10^{39}$ \junits})  &  \\ \hline 
Seyferts   &0.80 $(\pm 0.18)$ &0.20 \\
LINERS     &0.46 $(\pm 0.05)$ &0.11 \\
HII        &0.16 $(\pm 0.02)$ &0.04 \\
passive &0.17 $(\pm 0.02)$ &0.04 \\
\end{tabular}
\end{table}

\section{DISCUSSION \& SUMMARY} 
We have estimated the local galaxy X-ray LF 
and volume emissivity for different spectroscopic subsamples 
of galaxies: Seyferts, LINERS, star-forming and passive 
galaxies. This was done by combining the 
optical LF  derived from the Ho et al.
spectroscopic sample of nearby galaxies with the 
$L_x/L_o$ relation derived from the Fabbiano et al. 
 atlas of X-ray galaxies. 
The local emissivity is $j\approx 1.6\pm0.2\times 10^{39}$ \junits  
in the 0.2-4 keV band, translating to 
 a contribution of 40 per cent to the 
 extragalactic X-ray background, assuming no evolution. 
 Our result is consistent with that of 
 Soltan et al. (1997) who find  $j\sim 2 \times 10^{39}$ \junits 
 (0.2-4 keV) from a cross-correlation of bright galaxy catalogues 
 with the \rosat all-sky survey background maps. 
  In the hard 2-10 keV X-ray band, cross-correlations of local optical 
 and {\it IRAS} galaxy catalogues with {\it HEAO-1} and {\it Ginga}  data,
 yield similar results: Lahav et al. (1993), 
  Miyaji et al. (1994), Barcons et al. (1995) and Carrera et al. (1995)
 find local emissivities ranging from 
 $j\sim 1.0-1.7\pm0.6\times 10^{39}$ \junits, in the 0.2-4 keV 
 band, where we assumed for the conversion a spectral index of $\Gamma=1.7$. 
  Interestingly, cross-correlations of optical plates with 
 deep \rosat pointings yield lower emissivities  
  $j\sim 0.5\pm 0.07 \times 10^{39}$ \junits (0.2-4 keV) locally,
 (Roche et al. 1995, 1996, Almaini et al. 1997).  
 However, as the median redshift of the galaxies in Almaini et al. (1997)
  is z=0.45 while in our sample is z=0.004, cosmological 
 evolution may play an important role. 
  
 The largest contribution to our derived galaxy emissivity comes 
 from AGN ie. the Seyfert galaxies and LINERS. 
 Our Seyfert galaxy emissivity is in good 
 agreement with that of Barcons et al. (1995), 
 $j\sim 1.1\pm 0.2 \times 10^{39}$ \junits (0.2-4 keV), 
 derived on the basis 
 of IRAS $12\mu m$ and {\it HEAO-1} data.
 Unfortunately, due to the  
 small number statistics we are unable to distinguish between 
 different Seyfert sub-classes. 
 Note that, our Seyfert LF is in good agreement with the local QSO LF 
 of the RIXOS survey (Page et al. 1996),  at bright
 luminosities ($L_x \magcir 10^{42}$ \lunits)  
 where the Seyfert LF is dominated by the Seyfert 1 population (see Fig. 2). 
 At fainter luminosities our Seyfert LF is systematically
 above the RIXOS QSO LF. 
 A new result from our analysis is that LINERS 
 appear to contribute substantially  
 to the local volume emissivity.
  From Fig.2 it is evident that Seyferts dominate the 
 bright luminosities while LINERS play an important role 
 at luminosities $\mincir 10^{42}$ \lunits.
 Indeed, the cross-correlation of the \rosat all-sky survey 
 Bright Source Catalogue 
 with the Ho et al. sample  yields 45 coincidences within 
 1 arcmin (Zezas, Georgantopoulos \& Ward 1999). The majority of these 
 sources are LINERS (15) and Seyferts (13) while 4 objects 
 are intermediate between Seyferts and LINERS 
 according to the Ho et al. (1997a) classification. 
 The above result is marginally consistent with previous analyses:
 Miyaji et al. (1994) set up an upper limit to
 the non-Seyfert population (eg star-forming galaxies and LINERS) 
 emissivity of $j\approx 0.4\times 10^{39}$ \junits while
 Barcons et al. (1995) derive an $2\sigma$ upper limit  
 of $j\approx 0.7\times 10^{39}$ \junits. 
 This may be partly explained 
 by the fact that LINERS are mainly associated with early-type 
 galaxies (Ho et al. 1997b) which have low IR emission 
 and are thus under-represented  in {\it IRAS} galaxy samples. 
 The galaxy population which present no AGN activity such 
 as passive and star-forming galaxies, contribute only a small 
 fraction (4 per cent each class) of the XRB intensity 
 in the 0.2-4 keV band. 
 This is  somewhat higher than the 
  star-forming galaxy emissivity derived from the 
 the {\it IRAS} galaxy LF (Treyer et al. 1992). 

 In conclusion, the extragalactic X-ray light 
 appears to be dominated by accretion processes 
 rather than star-forming activity, in agreement with 
 previous predictions by Miyaji et al. (1994), Barcons et al. (1995). 
 Star-forming and passive galaxies 
 produce less than 10 per cent of the XRB intensity 
 in the 0.2-4 keV band assuming no evolution. 
 This independently suggests that the vast majority of 
 the galaxies detected in deep \rosat and \asca fields 
 are associated with AGN (Seyferts but also LINERS).
  Future large effective area missions 
 such as XEUS will be able to probe the faint tail of the 
 galaxy LF up to high redshifts and thus to detect large numbers 
 of non-AGN, normal galaxies in a similar fashion 
 to deep optical surveys. 

\section{Acknowledgments}
SB acknowledges the Greek Fellowship Foundation for a
postgraduate studentship. We thank the referee Dr. T. Miyaji as well as  
 Dr. G.C. Stewart for numerous useful comments and suggestions.

\end{document}